\documentstyle[twocolumn,prl,aps]{revtex}
\begin{document}
\title{Time-frequency transfer with quantum fields}
\author{Marc-Thierry Jaekel$^a$ and Serge Reynaud$^b$}
\address{$(a)$ Laboratoire de Physique Th\'{e}orique, CNRS, ENS, UPS,\\
24 rue Lhomond, F75231 Paris Cedex 05 France\\
$(b)$ Laboratoire Kastler Brossel, UPMC, ENS, CNRS,\\
4 place Jussieu, F75252 Paris Cedex 05 France}
\date{LPTENS 96/02}
\maketitle

\begin{abstract}
Clock synchronisation relies on time-frequency transfer procedures which
involve quantum fields. We use the conformal symmetry of such fields to
define as quantum operators the time and frequency exchanged in transfer
procedures and to describe their transformation under transformations to
inertial or accelerated frames. We show that the classical laws of
relativity are changed when brought in the framework of quantum theory.

\end{abstract}

\vspace{0.5cm}

As known since the advent of relativity theory\cite{Relativity1905}, time is
not an absolute notion. Positions in time of events occuring at different
locations in space have to be compared through clock synchronisation
procedures. From a practical point of view, electromagnetic signals are used
to transfer a time or frequency reference, allowing to compare respectively
the time delivered by remote clocks or their clock rates\cite{TimeFrequency}%
. The quality of these procedures is optimized by using a field pulse having
respectively a short duration or a good spectral purity. At some high level
of precision, synchronisation procedures have to reach a limit associated
with the quantum nature of the signals used in the transfer \cite{SW}. Time
and frequency of a field pulse cannot be defined simultaneously with an
arbitrary precision, so that clock synchronisation meets the problem of
localisability of quantum particles\cite{NW}. This leads to specific
difficulties for evaluating the spacetime dependent frequency shifts arising
from the effect of gravitational fields or non inertial frames. At the
quantum level, consistency of time and frequency transfers with frame
transformations is not ensured a priori by classical covariance. The aim of
the present letter is to build up such consistency properties by using the
underlying symmetries of quantum fields.

In 1905, Einstein proposed the hypothesis of light quanta\cite{Quantum1905}
and, a few months later, introduced the principle of relativity\cite
{Relativity1905}. In the latter paper, he incidentally noticed that energy
and frequency of an electromagnetic field change in the same manner in a
transformation from an inertial frame to another, thus implicitly referring
to the consistency of the hypothesis of light quanta with the principle of
relativity. Two years later, Einstein discovered that clock rates and
frequency shifts arising in transformations to accelerated frames depend on
position\cite{Relativity1907}. This allowed him to lay down the equivalence
of gravity and acceleration and to predict the existence of gravitational
redshifts. He again showed that energy and frequency change in the same
manner in the frame transformation.

In modern quantum theory, the similarity of energy and frequency changes has
to be interpreted as an invariance property for particle number. This
property is well known for Lorentz transformations, but it is usually not
admitted for transformations to accelerated frames. In particular, the
commonly considered hyperbolic parametrizations of accelerated frames
transform vacuum into a thermal bath\cite{Unruh}, spoiling the attempts to
accomodate the notions of particle number or vacuum to accelerated frames as
well as the efforts to understand the principle of equivalence in the
quantum domain\cite{Ginzburg}.

{}For Lorentz transformations, it is also usually considered that the
frequency changes, i.e. the Doppler shifts, are consistent with the
spacetime transformations described by the Poincar\'{e} group. Although
similar properties still hold for accelerated frames in general relativity,
they have not to our knowledge been brought in the framework of quantum
theory. In the absence of a satisfactory quantum description, one is left in
the quite uncomfortable position of having only the application of classical
covariance rules at one's disposal.

To be more explicit, let us consider a field used either as a frequency
reference or as a time reference to be shared by two remote observers. This
reference is in any case a quantity
preserved by field propagation. In classical physics, this quantity is the
field frequency $\omega $ or the light-cone variable $t-\frac xc$, where $t$
is the time coordinate, $x$ the space coordinate along which the transfer is
performed and $c$ the velocity of light. In order to build quantum
variables associated with the classical ones and to study how they
change in frame transformations, we will make use of symmetries of quantum
field theory.

Electromagnetism in four-dimensional spacetime is invariant not only under
Lorentz transformations, which fit inertial motions, but also under the
larger group of conformal transformations which fit uniformly accelerated
motions\cite{Conformal}. It follows that particle number, and hence vacuum,
have their definitions invariant under conformal transformations to
accelerated frames. In the present letter, we will use these invariance
properties to obtain a consistent quantum description of time and frequency
transfers and to describe how they are affected by transformations to
inertial or accelerated frames. We will present the discussion for the
simple case of a scalar massless field in two-dimensional spacetime. With
minor reservations to be indicated in the following, the results can be
translated to electromagnetic fields in four-dimensional spacetime, at the
price of more technical algebraic manipulations, but with their physical
significance preserved.

A free massless scalar field $\phi \left( t,x\right) $ in two-dimensional
spacetime is the sum of two counterpropagating components:
\begin{equation}
\phi \left( t,x\right) =\varphi ^{+}\left( t-x\right) +\varphi ^{-}\left(
t+x\right)
\end{equation}
{}From now on, we use natural spacetime units ($c=1$). In the following, we
study only one of the two counterpropagating components, that we simply
denote $\varphi \left( u\right) $:
\begin{eqnarray}
&&\varphi \left( u\right) =\int_0^\infty \frac{d\omega }{2\pi }\sqrt{\frac
\hbar {2\omega }}\left( a_\omega e^{-i\omega u}+a_\omega ^{\dagger
}e^{i\omega u}\right)  \nonumber \\
&&\left[ a_\omega ,a_{\omega ^{\prime }}^{\dagger }\right] =2\pi \delta
\left( \omega -\omega ^{\prime }\right)  \label{defphi}
\end{eqnarray}
$u$ is the light-cone variable to be shared in a time transfer procedure; in
the simple field theory considered here, $\omega $ represents the frequency
as well as the wavevector; $a_\omega $ and $a_\omega ^{\dagger }$ are the
standard annihilation and creation operators and $\delta $ is the Dirac
distribution.

The infinitesimal transformations of the light-cone variable $u$ are
characterized by relations between two coordinate systems $\overline{u}$ and
$u$ or, equivalently, by transformations of the field:
\begin{eqnarray}
\overline{u} &=&u+\varepsilon _mu^m\qquad \varphi \left( u\right) =\overline{%
\varphi }\left( \overline{u}\right)  \nonumber \\
\delta \varphi \left( u\right) &\equiv &\overline{\varphi }\left( u\right)
-\varphi \left( u\right) =-\varepsilon _mu^m\partial _u\varphi \left(
u\right)  \label{transfphi}
\end{eqnarray}
$\varepsilon _m$ is an infinitesimal quantity. All the conformal
transformations corresponding to any value of $m$ preserve the propagation
equation for a massless field theory in two-dimensional spacetime. Here, we
will consider only the transformations corresponding to $m=0,1,2$ which
respectively describe translations, dilatations and transformations to
accelerated frames. These transformations formally correspond to conformal
transformations in four-dimensional spacetime and they are known to preserve
vacuum fluctuations\cite{QSO95}. Other two-dimensional conformal
transformations, not corresponding to those in four-dimensional spacetime,
change vacuum fluctuations, which leads to the emission of radiation from
mirrors moving in vacuum with a non-uniform acceleration\cite{FDavies}.
Vacuum is invariant not only under the infinitesimal transformations (\ref
{transfphi}), but also under the finite transformations obtained by
exponentiation\cite{QSO95}, since such transformations form a group. This
latter property, which is essential for extending the symmetry associated
with the Poincar\'{e} group, is not obeyed by hyperbolic parametrizations of
accelerated frames. Vacuum is indeed invariant under infinitesimal
hyperbolic transformations, but transformed into a thermal state under
finite ones\cite{Unruh}.

In consistency with the canonical commutation relations (\ref{defphi}), the
field transformation (\ref{transfphi}) may be described by commutators with
conformal generators defined as moments of the stress tensor\cite{Itzykson}:
\begin{eqnarray}
\delta \varphi &=&\frac{\varepsilon _m}{i\hbar }\left[ T_m,\varphi \right]
\nonumber \\
T_m &=&\int u^me\left( u\right) du\qquad e\left( u\right) =:\left( \partial
_u\varphi \left( u\right) \right) ^2:  \label{defTm}
\end{eqnarray}
The symbol $:~:$ prescribes a normal ordering of products of operators, and
means that the generators vanish in vacuum. Vacuum invariance
under the action of $T_0$, $T_1$ and $T_2$ ensures the
consistency of this definition. The conformal generators are also conserved
quantities: the translation generator $T_0$ is the energy-momentum operator
associated with the light-cone variable $u$; $T_1$ corresponds to
dilatations and $T_2$ to transformations to accelerated frames. The
commutation relations between the conformal generators\cite{CAlg} are
recovered, either by inspecting the composition law for coordinate
transformations, or by evaluating quantum commutators:
\begin{equation}
\left[ T_m,T_n\right] =i\hbar \left( n-m\right) T_{m+n-1}\qquad m,n=0,1,2
\label{ConfAlg}
\end{equation}

In order to discuss the invariance of the photon number, we introduce
spectral decompositions for the conformal generators $T_0$ and $T_1$:
\begin{eqnarray}
T_m &=&\int_0^\infty \frac{d\omega }{2\pi }\rho _m[\omega ]  \nonumber \\
\rho _0[\omega ] &=&\hbar \omega n_\omega \qquad n_\omega =a_\omega
^{\dagger }a_\omega  \nonumber \\
\rho _1[\omega ] &=&\hbar \omega \sqrt{n_\omega }\left( \partial _\omega
\delta _\omega \right) \sqrt{n_\omega }\qquad a_\omega =e^{i\delta _\omega }%
\sqrt{n_\omega }  \label{defrho}
\end{eqnarray}
The density $\rho _0$ is related to the particle number density $n_\omega $
while the density $\rho _1$ is also related to the phase operators $\delta
_\omega $. Although the definition of these operators is ambiguous\cite{BP86}%
, the density $\rho _1$ is properly defined and it vanishes in vacuum.
Notice that the operators $\partial _\omega \delta _\omega $ are hermitian
even for non-hermitian definitions of the phases\cite{BJP}, and that they
are obtained by differentiating phases $\delta _\omega $ versus frequency $%
\omega $ in complete analogy with the semiclassical definition of scattering
phase-delays\cite{Wigner}. The transformation of the energy-momentum density
$\rho _0$ is easily obtained from the field transformation (\ref{transfphi}%
):
\begin{eqnarray}
\left[ T_0,\rho _0[\omega ]\right] &=&0  \nonumber \\
\left[ T_1,\rho _0[\omega ]\right] &=&i\hbar \omega \partial _\omega \rho
_0[\omega ]  \nonumber \\
\left[ T_2,\rho _0[\omega ]\right] &=&2i\hbar \omega \partial _\omega \rho
_1[\omega ]  \label{transfrho}
\end{eqnarray}
$\rho _0$ is unchanged under $T_0$, while it is changed under $T_1$ through
a mapping in the frequency domain equivalent to the Doppler shift. Its
transformation under $T_2$ is determined by the density $\rho _1$, thus
appearing as a spectral decomposition of the commutator $\left[
T_2,T_0\right] $ of equation (\ref{ConfAlg}). Equations (\ref{transfrho})
imply that the photon number $N$ is preserved under the three generators:
\begin{equation}
N=\int_0^\infty \frac{d\omega }{2\pi }n_\omega \qquad \left[ T_m,N\right] =0
\label{invarN}
\end{equation}
Hence, the action of $T_2$, like that of $T_1$, amounts to a redistribution
of particles in the frequency domain without any change of the total
particle number. In particular, vacuum, the $N=0$ state, is preserved under $%
T_2$. The transformations (\ref{transfrho}) have a simple interpretation
when the spacetime distribution of the field may be considered as
dispersionless. In this case, the operator $\partial _\omega \delta _\omega $
appearing in eqs. (\ref{defrho}) may be semiclassically approximated as the
classical light-cone variable $u$. The Doppler shift under $T_1$ is then
proportional to $\omega $, while the frequency shift under $T_2$ is
proportional to $2\omega u$, in consistency with the classical predictions%
\cite{Relativity1907}.

We will now write down expressions generalizing the classical transformation
laws to the quantum domain. To this aim, we first define operators $U$ and $%
\Omega $ associated with the light-cone variable $u$ and field frequency $%
\omega $. To emphasize the physical content of the equations, we will denote
$E$ the energy, which is also the translation generator, $D$ the dilatation
generator and $C$ the generator of conformal transformations to accelerated
frames:
\begin{equation}
E\equiv T_0\qquad D\equiv T_1\qquad C\equiv T_2  \label{defEDC}
\end{equation}
In the case of a scalar field considered here, or equivalently of spin-$0$
particles\cite{NW}, the operator $U$ is simply defined as the center of
inertia of field energy\cite{Relativity1906}, that is precisely for any
state orthogonal to vacuum ($E\neq 0$) as the ratio of $D$ and $E$ (compare
with eqs. (\ref{defTm})):
\begin{equation}
U=\frac 12\left\{ D,E^{-1}\right\}  \label{defU}
\end{equation}
We have taken care of the non-commutativity of the generators by
symmetrizing the expression; $\left\{ ~,~\right\} $ denotes an
anticommutator. The operator $\Omega $ is then defined as the ratio of the
energy $E$ to the particle number $N$:
\begin{equation}
\Omega =\frac E{\hbar N}  \label{defOm}
\end{equation}
{}For a single particle state ($N=1$), $\Omega $ plays exactly the same role
as $E=\hbar \Omega $. For a more general state however, the quantum
fluctuations of $\Omega $ and $E$ have independent meanings, since $N$ also
possesses its proper quantum fluctuations. $\Omega $ is thus a new quantum
concept which represents the mean frequency of the field quanta. Note that $%
N $ always commutes with $U$ as well as with $\Omega $:
\begin{equation}
\left[ N,U\right] =\left[ N,\Omega \right] =0
\end{equation}

Having given definitions for the operators $U$ and $\Omega $, we discuss in
the following how these definitions are affected by frame transformations.
To avoid any ambiguity, we may first recall that the light-cone variable $u$
and the frequency $\omega $ are preserved by field propagation in a
classical analysis. This property is still true in the present quantum
analysis, where the operators $U$ and $\Omega $ are conserved quantities,
which may thus be used to transfer time or frequency information between two
remote observers. However these operators are not invariant under frame
transformations. According to the discussion in the introduction, their
transformations are expected to reveal the basic relativistic properties of
time and frequency, within the framework of quantum theory.

The commutator $\left[ D,E\right] $ (see eqs (\ref{ConfAlg},\ref{defEDC})):
\begin{equation}
\left[ D,E\right] =-i\hbar E  \label{coDE}
\end{equation}
implies that the operator $U$ transforms under $E$ and $D$ as the classical
variable $u$ in the corresponding frame transformations\cite{NW}:
\begin{equation}
\left[ E,U\right] =i\hbar \qquad \left[ D,U\right] =i\hbar U  \label{transfU}
\end{equation}
This means in particular that $U$ is canonically conjugated to $E$. The
commutator $\left[C,E\right] $ then reads as:
\begin{equation}
\left[ C,E\right] =-2i\hbar D=-i\hbar \left\{ E,U\right\}  \label{coCE}
\end{equation}
The frequency shifts are finally given by equations (\ref{coDE},\ref{coCE})
combined with the invariance (\ref{invarN}) of $N$:
\begin{equation}
\left[ D,\Omega \right] =-i\hbar \Omega \qquad \left[ C,\Omega \right]
=-i\hbar \left\{ \Omega ,U\right\}  \label{transfOm}
\end{equation}
These laws reproduce the Doppler shifts associated with Lorentz
transformations as well as the position dependent frequency shifts arising
in transformations to accelerated frames. They fit the form of the
classical transformation laws\cite{Relativity1907}, while holding in any
quantum state orthogonal to vacuum. As already alluded to in the
introduction, the consistency between energy change and frequency change in
frame transformations reflects the invariance of the particle number.

These results bring the derivation of frequency shifts in the framework of
quantum theory. Precisely, frequency shifts may be evaluated from the
quantum transformation laws (\ref{transfOm}), which are identical to the
classical laws, but do not merely rely upon a classical covariance rule. A
fact of great interest for the physical analysis of time-frequency transfer
is that these expressions are available in the same theoretical framework
where quantum fluctuations of the various physical quantities may be
analyzed. They may thus be considered as setting the quantum limits in
time-frequency transfer. The canonical commutator $\left[ E,U\right] $ may
indeed be read:
\begin{equation}
\left[ \Omega ,U\right] =\frac iN
\end{equation}
In the limiting case of a large number of particles, this commutator goes to
$0$. This allows to build field pulses with nearly dispersionless
distributions of $\Omega $ and $U$ and to perform time-frequency transfer in
a semiclassical regime.

We have found that the transformations (\ref{transfU}) of the position
operator $U$ under the generators $E$ and $D$, as well as the
transformations (\ref{transfOm}) of the frequency operator $\Omega $ under
all generators have the simple form required by 'classical relativity'\cite
{Relativity1907}. We show now that the transformation of $U$ under $C$ does
not conform to these classical covariance rules. To this aim, we introduce
the following quadratic form $\Delta ^2$ of the generators, which is a Casimir
invariant of the conformal algebra (\ref{ConfAlg}):
\begin{eqnarray}
\Delta ^2 &=&\frac 12\left\{ C,E\right\} -D^2  \nonumber \\
\left[ E,\Delta ^2\right] &=&\left[ D,\Delta ^2\right] =\left[ C,\Delta
^2\right] =0  \label{defQ}
\end{eqnarray}
Definitions (\ref{defU},\ref{defQ}) and commutation relations (\ref{ConfAlg}%
) of the conformal algebra allow to rewrite the generator $C$ and its action
on the operator $U$ as:
\begin{eqnarray}
C &=&UEU+\frac{\Delta ^2}E+\frac{\hbar ^2}{4E}  \nonumber \\
\left[ C,U\right] &=&i\hbar \left( U^2-\frac{\Delta ^2}{E^2}-\frac{\hbar ^2}{%
4E^2}\right)  \label{defDelta}
\end{eqnarray}
The first term in each expression corresponds to the symmetric ordering of
their classical analogs. The other terms are corrections associated with the
pulse duration, as it follows from the relation:
\begin{equation}
\frac 12\int du\left\{ \left( U-u\right) ^2,e\left( u\right) \right\} ={%
\frac 1E}\left( \Delta ^2+{\frac{\hbar ^2}4}\right)
\end{equation}
It can be shown that $\Delta ^2$ has a non negative mean value in any field
state, and that it vanishes in any $1$-particle state. It may in principle
be made close to $0$ either by using a very short pulse or by using $1$%
-particle pulse. In contrast, the terms proportional to $\hbar ^2$ appear as
purely quantum corrections to the classical terms. They only
become negligible at the semiclassical limit where a large number of
particles is used ($N\gg 1$).

The transformation (\ref{defDelta}) for the operator $U$ under the
acceleration generator $C$ differs from its classical covariant analog.
Corrections however involve operators which commute with $E$, and are
therefore unchanged if the pulse used for the transfer is delayed. It may be
stated equivalently that time transfer procedures are found to be invariant
under time translation. This statement is the expression of the consistency
between time and frequency transfers in the quantum domain. It does however
not imply that the corrections still disappear when successive transfer
operations are performed. In this case, a sequence of field pulses has
indeed to be used and the corrections may vary from one pulse to the next
one.

The problem raised here is not a practical limitation in present
time-frequency metrology, even at the state-of-the-art level. As a matter of
principle however, we emphasize once more that clock synchronisation has to
involve quantum fields. As a consequence, the consistency of synchronisation
operations, which follows in the quantum domain from conformal symmetry,
entails a departure from the classical covariant laws for frame
transformations.

The common conception of spacetime associated with the theory of general
relativity is known to remain ambiguous\cite{Norton}.
The result of the present letter extends the connection between symmetries of
quantum
fields and relativitistic properties of spacetime from inertial frames
to accelerated frames. It thus advocates
a novel conception of spacetime which would be free from its difficulties
inherited from classical physics\cite{Rovelli}.

\end{document}